\documentstyle[prb,aps]{revtex}

\begin{document}



\input epsf.sty
\twocolumn[\hsize\textwidth\columnwidth\hsize\csname %
@twocolumnfalse\endcsname
 
\draft

\widetext


\title{Finite-size effects in layered magnetic systems}

\author{Dragi Karevski and Malte Henkel}
\address{Laboratoire de Physique du Solide\cite{URA},
Universit\'e Henri Poincar\'e Nancy I, B.P. 239, \\
F - 54506 Vand{\oe}uvre l\`es Nancy Cedex, France}

\maketitle
  
\begin{abstract}

Thermal and magnetic effects in a system consisting of thin layers of coupled
Ising spins with $S=\frac{1}{2}$ and $S=1$ are considered. 
The specific heat and the correlation length display maxima at two different
temperatures. It is discussed in what sense these maxima can be interpreted as 
a finite-size rounding of a thermodynamic singularity associated with a 
phase transition. The connection with ordinary, extraordinary and special
surface phase transitions is made. In $2D$, the surface critical exponents are 
calculated from conformal invariance. The bulk and surface finite-size scaling 
of the order parameter profiles at the transition points is discussed. 
In $2D$, an exact scaling function for the profiles is suggested through 
conformal invariance arguments for the (extra)ordinary transition.
\end{abstract}

\pacs{PACS numbers: 05.50+q, 75.40-s, 68.35Rh, 11.25Hf }

\phantom{.}
]

\narrowtext
\section{Introduction}

Considerable effort has been recently devoted to the understanding
of magnetic thin films. The behaviour of 
magnetic insulators such as the transition-metal difluorides 
can be described in terms of short-ranged interaction
models which makes the comparison with theory considerably 
simpler \cite{Lede93a}. In particular, these materials can be
epitaxially grown in very thin films to the point that specific theoretical
concepts such as finite-size scaling close to a second order critical
point \cite{Barb83,Priv90} become experimentally verifiable. 
Such a study was performed \cite{Lede93a}
for the (FeF$_2$)$_n$(ZnF$_2$)$_m$ superlattice, 
where the magnetic interactions
within a single FeF$_2$ layer can be described 
in terms of a spin $S=2$ bcc Ising
model (with the different FeF$_2$ layers sufficiently far apart that 
free boundary conditions can be assumed for each of them). 
The data for the thermal expansion coefficient $\alpha(T)$,
which is experimentally observed \cite{Lede93a} to be proportional to the 
magnetic contribution to the
specific heat, show finite-size shifts of the critical point and 
rounding of the thermodynamic singularity in quantitative
agreement with finite-size scaling theory \cite{Lede93a}. 
Subsequently, the thermal properties
of (FeF$_2$)$_n$ (CoF$_2$)$_n$ superlattices, 
where two different magnetic layers
interact, were studied \cite{Lede93}. The thermal expansion coefficient
was studied as a function of temperature and of the layer 
thickness $n$. For $n$ small, $\alpha(T)$ was found to show a single maximum,
while for $n$ larger, two maxima of $\alpha(T)$ as a funtion of temperature
were observed \cite{Lede93}. Besides studying thermal properties, 
it is also possible to
explore experimentally the magnetic properties of single monolayers
through M\"o{\ss}bauer spectroscopy \cite{Baue95} and to investigate the
resulting order parameter profiles. 

In an attempt to provide a theoretical description of these layered
magnetic systems 
beyond mean-field theory, we consider here as a simple toy model two coupled
magnetic subsystems, where each subsystem contains $n$ parallel layers 
of classical Ising spins, with $S=\frac{1}{2}$ and $S=1$ respectively. 
We assume nearest
neighbor couplings between the spins. For simplicity, we also assume that
the coupling between spins within a layer is independent of $S$. 
We cannot expect with such a simple model to reproduce quantitatively
any of the experiments mentioned above, but we shall use our model as a
means to obtain and to test scaling descriptions of the experimentally observed
phenomena. Scaling should apply to more realistic situations. 
We shall thus write down the model in the form best suited for
numerical treatment. 

Two main simplifications are employed. 
First, we work in {\em two dimensions},
considering layers of spin chains rather than the three-dimensional layers
of films studied experimentally. We expect that the scaling
picture used to describe the critical behaviour can be applied in two as well
as in three dimensions. Second, an extreme anisotropic limit is 
used \cite{Suzu71}, where coupling constants
between different layers are becoming very small while within a layer they
become large. Then the task of calculating the thermal behaviour of the system
amounts to studying the ground state properties of 
the quantum Hamiltonian 
\begin{eqnarray}
H &=& -\frac{1}{2\zeta} \left[
\sum_{\ell=1}^{n-1}  \sigma_{\ell}^z \sigma_{\ell+1}^z  
+\sum_{\ell=n+1}^{2n-1} \kappa S_{\ell}^z S_{\ell+1}^z  
\right.\nonumber \\
&+& \left. t \sum_{\ell=1}^n \left( \sigma_{\ell}^x + S_{\ell+n}^x \right)+
\gamma 
\left( \sigma_n^z S_{n+1}^z + \sigma_{1}^z S_{2n}^z \right) \right] 
\label{HamDef}
\end{eqnarray}
where $\sigma^{x,z}$ are the spin $\frac{1}{2}$ Pauli matrices 
and $S^{x,z}$ are 
spin $1$ matrices. It is well known \cite{Suzu71,Chri93} 
that the critical behaviour of this quantum chain is in
the same universality class as the two-dimensional model of classical
Ising spins described above (experimentally, this correspondence has 
recently been demonstrated \cite{Bitk96} for the dipolar-coupled
$3D$ Ising ferromagnet LiHoF$_4$), 
but the numerical treatment of $H$
is considerably easier than the corresponding calculation in the classical
spin model using the transfer matrix. The critical behaviour of several
coupled Ising systems with spin $\frac{1}{2}$ in all subsystems had been 
investigated earlier \cite{Hinr90,Berc91,Zhan96}. 

Let us explain the terms arising in $H$ by making the analogy
with the two-dimensional model of classical Ising spins.  
The terms $\sigma_{\ell}^z\sigma_{\ell+1}^z$ describe the 
interactions between spins in different layers and the terms
$\sigma_{\ell}^x$ describe the interactions within a single spin layer 
(and similarly for the $S^{x,z}$). 
The coupling $t$ plays the role of a temperature. The $S$-independence
of the transverse field $t$ reflects our assumption that the spin-spin coupling
within a layer is spin-independent. Finally, $\gamma$
describes the coupling between the two subsystems. The spatial coordinate
$\ell$ corresponds to the direction perpendicular to the magnetic layers. 

The following symmetries of $H$ are immediate. First, $H$ is invariant under
the global spin reversal 
$\sigma^z \rightarrow -\sigma^z, S^z\rightarrow -S^z, 
\sigma^x\rightarrow\sigma^x,
S^x\rightarrow S^x$. 
Those states which are invariant under this transformation are
said to be {\em even}, all other states are said to be {\em odd}. Thus $H$
is block-diagonalized into an even and an odd sector. Second, the spectrum of
$H$ is independent of the sign of $\gamma$, because $H(\gamma)$ is changed
into $H(-\gamma)$ through the similarity transformation 
$\sigma^z\rightarrow -\sigma^z,
S^z \rightarrow S^z$.

For each subsystem alone,
that is for $\gamma=0$ and $n\rightarrow\infty$, 
there is a critical point at $t=t_{c,S}$ with \cite{Pfeu70,Hofs96}
\begin{equation} \label{critpt}
t_{c,\frac{1}{2}} = 1 \;\; , \;\; t_{c,1}/\kappa = 1.32587(1)
\end{equation}
respectively. Varying $\kappa$ thus allows to change the ratio 
between the critical points
$t_{c,S}$ in the $S=\frac{1}{2},1$ systems. Finally, $\zeta$ is
a normalization constant which will be needed below 
in connection with the conformal
invariance description of the spectrum of $H$ at criticality. 

We are interested in the following observables which will be studied through
their quantum analogues.\cite{Suzu71,Chri93} The free energy of the
two-dimensional classical spin model corresponds to the ground state
energy $E_0$ of $H$. Similarly, 
thermal averages $<X>$ correspond to ground
state expectation values $\langle 0|X|0\rangle$. We also need 
the characteristic lengths
$\xi_{1,2}$ of the spin-spin and energy-energy correlations
(for $r\rightarrow\infty$)
\begin{eqnarray}
G_{\sigma}(r) = <\sigma(r) \sigma(0)> - <\sigma(r)><\sigma(0)> &\sim& 
e^{-r/\xi_{1}}
\nonumber\\
G_{\epsilon}(r) = <\epsilon(r) \epsilon(0)> - <\epsilon(r)><\epsilon(0)> 
&\sim& e^{-r/\xi_{2}}
\end{eqnarray} 
where $r$ is parallel to the individual layers. One has 
$\xi_{1,2}^{-1}=E_{1,2} - E_0$ where $E_{1,2}$ are the 
energies of the first excited states in the odd and the even sector, 
respectively.

The numerical technique used is completely standard, see Refs. 
\onlinecite{Dago94,Chri93} for details. We use the L\'anczos
algorithm to find the first few lowest eigenvalues of $H$ and the corresponding
eigenvectors. Finite-size scaling is then used to obtain estimates for
the critical quantities which are then numerically extrapolated for 
$n\rightarrow\infty$.

This paper is organized as follows. In section 2, we discuss the phase diagram
and comment on a subtlety in finite-size scaling. Section 3 describes
the calculation of the surface critical exponents in $2D$ through conformal
invariance techniques. In section 4, we
present our results for the order parameter profiles. Finally, we give our
conclusion in section 5.

\section{The phase diagrams}

Our starting point is the experimental observation \cite{Lede93} that
the specific heat $C$ as a function of the temperature will show one
or two maxima depending on the thickness $n$. We therefore begin, 
with a consideration
of this quantity. However, the explicit calculation of the
second derivative $-t \partial^2 E_0(t)/\partial t^2$ of the free energy is 
cumbersome. To avoid this, recall the fluctuation-dissipation relation 
$C \sim \sum_{\vec{r}} G_{\epsilon}(\vec{r}\,)$ 
together with the scaling form, which should
be valid near criticality 
\begin{equation} \label{escal}
G_{\epsilon}(r) = r^{-2 x_{\epsilon}} g(r/\xi_2)  
\end{equation}
where $x_{\epsilon}=(1-\alpha)/\nu$ and $\alpha,\nu$ are conventional
critical exponents. Then, up to nonsingular 
background terms, the relation 
\begin{equation}
C(t) \sim (\xi_2(t))^{\alpha/\nu}
\end{equation}
should hold \cite{alnull}. 
Since we are here only interested in the leading critical
behaviour, it is sufficient for us to consider the {\em second} gap 
$\xi_{2}^{-1} = E_2 -E_0$ of $H$. The scaling behaviour of $\xi_2$ is simply
related to the scaling of the specific heat $C$ and moreover the
temperature dependence not too far away from the critical region of both
$\xi_2(t)$ and $C(t)$ should be qualitatively similar. Finally,
$\xi_2$ is readily calculated through
the L\'anczos algorithm \cite{Dago94,Chri93}.  
We point out that the spin correlation length $\xi_1$
does {\em not} enter into the scaling form (\ref{escal}), because it couples
to quantities which are odd under spin reversal while $G_{\epsilon}$ is 
even \cite{xis}. 

\begin{figure}
\centerline{\epsfxsize=3.25in\epsfbox
{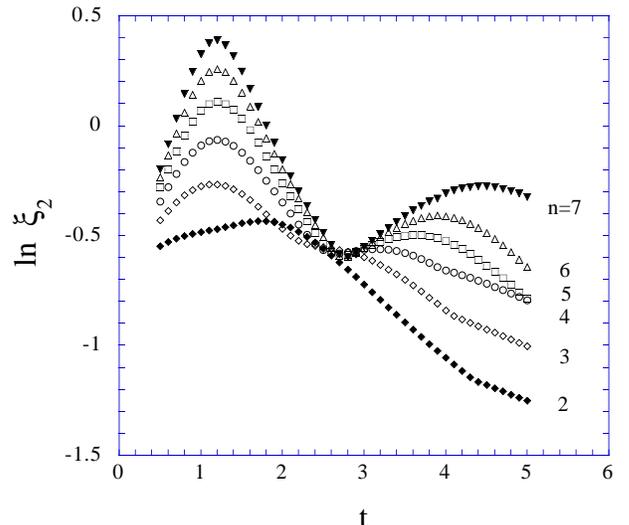}}
\caption{The energy correlation length $\xi_{2}$ as a function of $t$ and
different layer thicknesses $n$ for $\kappa=4$ et $\gamma=1$. The size of
the layers is $n=2\ldots 7$ from bottom to top. \label{fig1}
}
\end{figure}

In figure~\ref{fig1} we show $\ln \xi_2$ as a function of 
$t$ for different layer thicknesses $n$. 
We observe that for a very thin layer ($n=2,3$), there is 
only a single maximum present while two maxima develop for larger values of
$n$. Comparing the location of the maxima for $n$ finite with the
known values from (\ref{critpt}) for their $n\rightarrow\infty$ limit, 
we see that
the shift in the effective critical temperatures are quite large. 
Both maxima appear to show a systematic build-up normally considered 
typical of a thermodynamic singularity rounded by finite-size 
effects \cite{Barb83}. These observations, of one or two maxima depending
on the value of $n$, 
large finite-size shifts of the pseudo critical temperatures and
a rounding of the thermodynamic singularity, are in qualitative agreement with
experiment \cite{Lede93,Ramo90}. 

\begin{figure}
\centerline{\epsfxsize=3.25in\epsfbox
{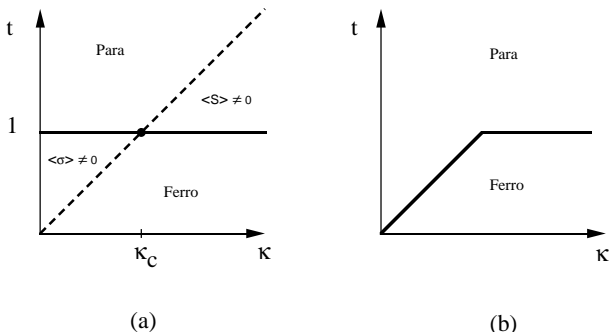}}
\caption{Phase diagrams for the two thermodynamic limits, for
two-dimensional systems. (a) Two layers
with  $n$ spins each, periodic boundary conditions and $n\rightarrow\infty$. 
(b) Superlattice of $m$ bilayers of $n$ spins of each kind and 
$m\rightarrow\infty$. \label{fig2}
}
\end{figure}

Before we can make this conclusion however, one should realize that the
models usually considered in theoretical calculations and the superlattices
studied experimentally are different. We shall refer to these as case A and
case B, respectively. These cases differ in the way one goes from the finite
system to an infinite one and it is only for the infinite system where
a true phase transition can occur. Consequently, the phase diagrams for
cases A and B are different. (We reemphasize that in the following discussion
we refer to two-dimensional systems while experiments are carried out in three
dimensions.) 
\begin{enumerate}
\item[A)] We take a layer of $n$ spins $\frac{1}{2}$ and a layer of 
$n$ spins $1$. Periodic boundary conditions as explicitly written 
in (\ref{HamDef}) are used. The phase diagram which results when
$n\rightarrow\infty$ is given in figure~\ref{fig2}a. 
\item[B)] In experiments \cite{Lede93,Lede93a,Baue95}, 
a procedure analogous to the following is used. 
One takes $n$ spins $\frac{1}{2}$ and $n$ spins $1$ and 
repeats this double
layer $m$ times. Typically, $n$ is small and fixed but $m$ is large and 
formally, one should take a limit
$m\rightarrow\infty$.  This leads to the phase diagram in figure~\ref{fig2}b.
\end{enumerate} 

We first consider case A. Then, 
each of the two subsystems can develop long-range order by itself. 
Consequently, the phase diagram (figure~\ref{fig2}a) will show four
different phases. There is a paramagnetic phase where the whole system is
disordered, two distinct phases where either the spin $\frac{1}{2}$ variables
($<\sigma>\neq 0$) or the spin $1$ variables 
($<S>\neq 0$) are ordered while the
other subsystem is disordered and a ferromagnetic phase where the system is
fully ordered. The transition lines are given by eq.~(\ref{critpt}) (full
line for $t_{c,\frac{1}{2}}$ and dashed line for $t_{c,1}$). In this
case, the maxima observed in figure~\ref{fig1} should be interpreted as
true thermodynamic singularities rounded by finite-size effects. 
Since we shall below concentrate on the 
properties of the order parameter close
to the subsystem boundaries, we label 
these transitions by its surface critical
properties, following the theory of
surface phase transitions \cite{Dieh86}. For the transitions from
the paramagnetic phase to one of the partially ordered phases, one of 
the subsystems is still
disordered and the order parameter of that subsystem
which undergoes ordering will
vanish at the boundary between the two subsystems. Along this line we
have an {\em ordinary} transition.  On the other hand,
for the transitions from the partially ordered phases
to the ferromagnetic phase one 
subsystem is already ordered which fixes the order parameter
of the other subsystem at the subsystem boundary. Here we have 
an {\em extraordinary} transition. At the meeting point of
the transition lines there is a {\em special} transition \cite{zweidrei}. 
The scaling of the order parameter close to the subsystem boundaries
is described by a different exponent than for the bulk, 
see Ref. \onlinecite{Dieh86}. 
These local critical exponents are in $2D$ readily calculated using conformal 
invariance techniques, see section 3.

For case B, 
corresponding to figure~\ref{fig2}b however, the situation is different. 
Since each of the subsystems only contains a finite number of layers,
the superlattice can only order as a whole. Thus the phase diagram contains
a paramagnetic phase and a single ordered ferromagnetic phase. If $t$
is sufficiently small, a layer of $n$ spins may act as a giant
spin and produce a strong thermal signal leading to a {\em finite} maximum
of the specific heat or of related quantities. 
Since $n$ is finite, however, there is no long-range order and the magnetic
moments of each layer are independent of each other. Then the specific
heat as a function of temperature will show two peaks, but only the one
at {\em lower} temperatures will then correspond to a (shifted and rounded)
phase transition and will develop a true singularity as $m\rightarrow\infty$. 
Working in the framework of case B, it is misleading to call the location
of the larger temperature maximum a (pseudo) critical point.

From now on we always consider case A. Then, both maxima in $\xi_2$ can
be interpreted as signalling a transition. 
Also, we shall perform the subsequent
scaling analyses just for the two-dimensional system, since the changes
which might be needed in three dimensions are immediate and discussed in
detail in the literature \cite{Barb83,Priv90}.
 
\begin{figure}
\centerline{\epsfxsize=3.25in\epsfbox
{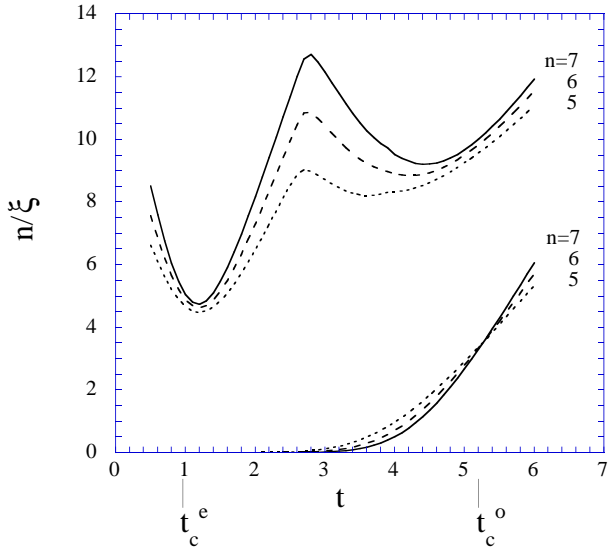}}
\caption{Scaled inverse correlation lengths $n/\xi_{1,2}(t)$ as a function of
$t$, for $\kappa=4$ and $\gamma=1$ and several layer thicknesses $n$. 
The critical points are
labelled $t_c^o$ and $t_c^e$ corresponding to the ordinary and extraordinary
transitions. The lower (upper) curves correspond to $\xi_1$ ($\xi_2$).
\label{fig3}
}
\end{figure}

How can one find the critical points
from the finite-lattice data, when the Hamiltonians are more complicated and
precise information on their location such as (\ref{critpt}) is not available
{\em a priori ?} 
Practically, the transition points are located using phenomenological
renormalization as derived from finite-size scaling \cite{Barb83,Priv90}. 
Consider the quantity $R(t;n) = n/\xi(t;n)$. Then finite-size estimates for
the critical point $t_c$ can be found by solving for $t$ the equation
\begin{equation} \label{eqfss}
R(t;n) = R(t;n+1) 
\end{equation}
if $n$ sufficiently large. A final value for $t_c$ is then obtained by
extrapolating the resulting sequence for $n\rightarrow\infty$. Carrying out
this procedure, a further subtlety is encountered as illustrated in 
figure~\ref{fig3}. 
Considering the finite-size scaling of the spin correlation length $\xi_1$
(lower curve), we see that the curves $R_1(t;n)$ intersect close to the
critical point $t_c^o$. This is the conventional behaviour found e.g. in
simple Ising models \cite{Barb83,Priv90}. 
For smaller values of $t$, $\xi_1^{-1}$ vanishes exponentially fast with $n$,
which reflects the ordering of the $S=1$ subsystem in this case. Thus in order
to find $t_c^e$, the second gap $\xi_2^{-1}$ 
is the natural quantity to look at and in fact represents in the partially
ordered phases the lowest physical excitation, just as $\xi_1^{-1}$ does in the
paramagnetic phase. 
Nevertheless, the curves $R_2(t;n)$ go through a
minimum close to $t_c$ 
and will eventually touch each other in the $n\rightarrow\infty$ limit,
but do not intersect. Although this is {\em not} in contradiction with the
theory of finite-size scaling (at $t=t_c$, (\ref{eqfss}) is
strictly valid for $n\rightarrow\infty$ only),
it is remarkable that at this point the conventional finite-size techniques
are no longer applicable. In order to get an estimate of $t_c$, one has to 
rely on locating a minimum of
$R_2(t)$ or some other criterion. We stress that at $t=t_c^e$ is the
{\em only} phase transition occuring in the model for case B. 

This type of behaviour should be generic and
although the example given does suggest that finite-size techniques may be
fruitfully employed in analysing experimental data, it also shows that 
some care may be required. We shall see in the next section that in spite
of the slightly unusual finite-size scaling, the spectrum of $H$ at all these
critical points 
is in full agreement with the conformal invariance predictions. 

To illustrate to what extent quantitative 
information about the critical behaviour can
be extracted from our still relatively small systems ($n\leq 7$), we consider
the determination of the correlation length exponent $\nu$. 
We look at the local maxima $t_{max}(n)$ 
of $\xi_2$ near to the point $t=t_c^o$. Finite-size scaling predicts 
\cite{Barb83,Priv90} that the temperature shift
\begin{equation} \label{DelT}
\Delta t (n) =  t_c^o- t_{max}(n)  \sim n^{-1/\nu} \;\; , \;\; 
n\rightarrow\infty
\end{equation}
\begin{figure}
\centerline{\epsfxsize=3.25in\epsfbox
{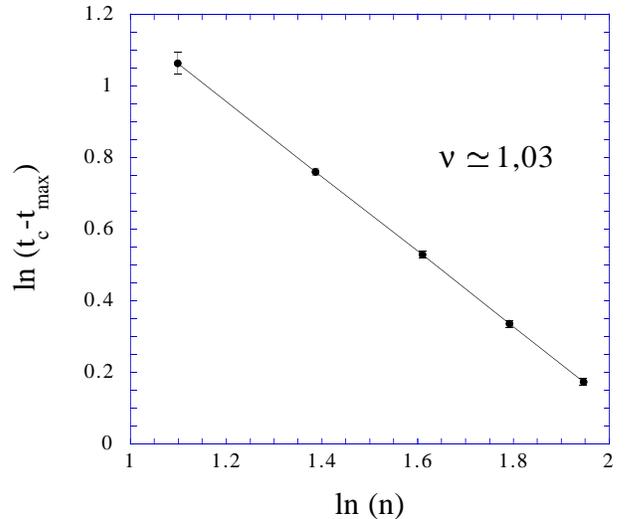}}
\caption{Determination of $\nu$ from the finite-size scaling of $t_c(n)$
near the ordinary transition for $\kappa=4$. \label{fig4}
}
\end{figure}
In figure~\ref{fig4}, we show a log-log plot of $\Delta t(n)$ vs. $n$ and find
that the asymptotic behaviour (\ref{DelT}) 
is already realised for small $n$ in our toy 
model, although the $t_c(n)$ are not at all close to the 
$n\rightarrow\infty$ value $t_c^o$, see figure~\ref{fig3}. From the slope
in figure~\ref{fig4}, we read off $\nu\simeq 1.03$, 
in good agreement with the exact result $\nu=1$. We point out that
also ($3D$) data for the thermal expansion coefficient 
for FeF$_2$/ZnF$_2$ superlattices
were successfully analysed this way \cite{Lede93a}, 
although also in this case $\Delta t(n)$
is large \cite{Ramo90}, leading to $\nu=0.64(4)$ in agreement with
the theoretical value $\nu\simeq 0.63$ for the $3D$ Ising model. 

\section{Critical exponents and conformal invariance}

We now describe the calculation of the critical exponents. Since we work with
a two-dimensional classical spin model universality class, we can use
conformal invariance techniques \cite{Chri93,Card87,Itzy89,Iglo93} 
for that purpose. Here the use of the quantum Hamiltonian rather than
the classical spin model becomes advantageous, since the surface critical
exponents which describe the local scaling close to the boundary between
the two subsystems and in which we are mainly interested 
are easily obtained from the low-lying excitation spectrum of $H$. We can
thus avoid the cumbersome procedure of calculating first an average 
$<X>$ and then subtracting from it its bulk contribution $X_b$
in order to get the surface term $X_s = <X> - X_b$ and then
analyse its scaling behaviour. In the next two subsections, we shall first
briefly collect the necessary background knowledge and shall then apply it
to the problem at hand.  

\subsection{Ordinary and extraordinary transitions}

In two dimensions (and consequently also for quantum chains) conformal
invariance specifies completely the scaling 
dimensions of all local observables.
For a given model the first few exponents are very easily
identified from the spectrum of the Hamiltonian $H$. For free or fixed
boundary conditions one has \cite{Card84} 
\begin{equation} \label{ConfInv}
\xi_i^{-1} = E_i - E_0 = n^{-1} \pi x_i
\end{equation}
where the exponents $x_i$ are the {\em local} critical exponents which
describe scaling near the boundary between the two subsystems and the index
$i$ labels the various scaling fields which occur in the model (usually, 
$i=1$ corresponds to the order parameter and $i=2$ to the energy density and
higher gaps correspond to the scaling fields which generate 
correction-to-scaling terms). The scaling of the gaps (\ref{ConfInv}) 
goes with $n$ and 
{\em not} with $L$ because only half of the system is critical at either the
ordinary or extraordinary transitions. 
However, in order to be able to apply (\ref{ConfInv}) to the spectrum of a
quantum chain such as (\ref{HamDef}), the normalization of $H$ must be
chosen such that energies and momenta are measured in the same units. One
way of doing this is to recall that the surface scaling dimension 
of the energy density
\begin{equation}
x_{\epsilon,s} =2
\end{equation}
which fixes the normalization $\zeta$. Furthermore, once the normalization
is fixed accordingly, the conformal algebra
acts as a dynamical symmetry which determines the spectrum of $H$ at
criticality, viz.
\begin{equation} \label{HVira}
E_i - E_0 =  \frac{\pi}{n}  L_0  + o(n^{-1})
\end{equation}
with $L_0$ being one of the generators of the Virasoro algebra
\begin{equation}
[L_j, L_k] = (j-k)L_{j+k}+\frac{c}{12}(j^3-j)\delta_{j+k,0}
\end{equation}
The universality class is
determined by the value of the {\em central charge} $c$, for the $2D$ 
Ising model \cite{Chri93,Card87,Itzy89} $c=\frac{1}{2}$. 

Furthermore, the spectrum of $H$ at criticality can be 
found from the representations of the Virasoro algebra, see Refs.
\onlinecite{Card87,Itzy89,Chri93} for details. These representations
are built from a highest weight state $|\Delta\rangle$ which is defined
through
\begin{equation}
L_0 |\Delta\rangle = \Delta |\Delta\rangle \;\; , \;\;
L_j |\Delta\rangle = 0 \;\; \mbox{\rm if $j>0$} 
\end{equation}
and acts as the ground state for a certain representation. Excited states
are generated by acting with the $L_{-j}$ ($j>0$) on $|\Delta\rangle$. Now, the
principle of unitarity of the underlying field theory restricts through
the Kac formula the possible values of $c$ and for each value of $c$ only
permits a finite number of possible values of $\Delta$. For $c=\frac{1}{2}$,
the only possible values are $\Delta=0,\frac{1}{16},\frac{1}{2}$. This leads
to the three unitary irreducible representations $(0)$, $(1/16)$ and $(1/2)$
of the $c=\frac{1}{2}$ Virasoro algebra. Now, the spectrum of $H$ for the
ordinary and the extraordinary transitions is given by \cite{Burk87,Berc91}
\begin{eqnarray}
H^{(o)} &=& (0) + (1/2) \qquad \mbox{\rm ordinary} \nonumber \\ 
H^{(e)} &=& 2 (0) \qquad \qquad \quad \mbox{\rm extraordinary}  
\end{eqnarray}
where the trivial prefactor $\pi/n$ is suppressed. The factor 2
for the spectrum at the extraordinary transition means that each level has
the double degeneracy of the representation $(0)$. Combining these predictions
with the formula (\ref{ConfInv}) for the energy gaps, the critical exponents
$x_i$ can be read off.

These predictions, which had already been 
checked for the spin $\frac{1}{2}$ before
\cite{Burk87,Berc91}, are fully reproduced in our model, in agreement with
the expected universality. As an example, we take
$\kappa=4$ and $\gamma=1$, but the results for the exponents do not
depend on these parameters. The values for 
$t$ at the ordinary and extraordinary 
transition are from eq.~(\ref{critpt}) $t_c^o=5.30348(4)$ and $t_c^e=1$. The
energy gap which is related through (\ref{ConfInv}) 
to the exponent $x_{\epsilon,s}$
is the lowest gap in the even sector. Lattices with up to $n=7$ were used. 
After extrapolation \cite{Chri93,Henk88}, we find $\zeta^{(o)}=2.319(6)$ and
$\zeta^{(e)}=0.996(5)$ for the ordinary and the extraordinary transitions, 
respectively \cite{norm}. In table~\ref{tab1}, 
we give the extrapolated estimates for the
first four rescaled gaps $x_i = (E_i-E_0)n/(\zeta\pi)$ 
for the ordinary and extraordinary transitions together with the predictions
following from conformal invariance. For the extraordinary transition, all
levels were found to be doubly degenerate in the $n\rightarrow\infty$ 
limit but with
an exponentially small splitting between pairs of levels for $n$ finite. 
As should be expected from the algebraic construction of the spectrum of
$H$, the differences between values of the $x_i$ which belong to the same
representation are integers. This reconfirms our determination of $\zeta$. 
For the ordinary transition, we see that $x_{\epsilon,s}=x_3=2$ and 
$x_1=x_{\sigma,s}=\beta_1/\nu=1/2$ is the surface magnetization exponent, 
where $\beta_1$ describes the scaling of the order parameter at the
surface, $m_1 \sim (T_c-T)^{\beta_1}$ and $\nu$ is the bulk correlation 
length exponent \cite{Dieh86}.

For the extraordinary transition, a little care is necessary in identifying
the surface exponents. The order parameter is odd under spin reversal and
the most local of all scaling operators. When this operator acts on the 
ground state, it creates the state with the lowest gap in the spectrum. Thus in
our model $x_{\sigma,s}^{(1)}=x_0=0$. On the other hand, in the literature 
(e.g. Ref. \onlinecite{Dieh86} and refs. therein), 
the extraordinary transition is
defined with respect to those degrees of freedom which become critical in the
presence of a boundary which is already ordered. To read off the corresponding
exponents, one should discard the double degeneracy of the spectrum, which
is merely due to the ordering of the other subsystem. We then have 
$x_{\epsilon,s}=x_{\sigma,s}^{(2)}=x_1=2$. This is in agreement
with the expected scaling 
relation \cite{Bray77,Dieh86} $x_{\sigma,s}=\beta_1^{ex}/\nu=2-\alpha=2$.

\subsection{Special transition}  

At the special transition, both subsystems become critical simultaneously. 
In addition, the Ising quantum chain has the peculiarity that the boundary
coupling $\gamma$ is marginal. 
The critical behaviour of the model can be described
using previous results for the scaling behaviour of an Ising model with
(semi-)infinite defect lines, which has been extensively studied 
for a long time
\cite{Hinr90,Berc91,Zhan96,Bari79,Turb85,Henk87,Henk89,Oshi96}, 
see Ref. \onlinecite{Iglo93} for a review. The local critical exponents depend
continuously on the coupling $\gamma$. The mapping of coupled Ising layers to
a $2D$ Ising model with a star-like configuration of semi-infinite defect
lines was exactly derived for coupled spin $\frac{1}{2}$ Ising 
models \cite{Hinr90,Berc91,Zhan96}. The surface critical
exponents can be read off the energy spectrum \cite{Turb85}
\begin{equation} \label{turban}
\xi_{i}^{-1} = E_i - E_0 = L^{-1} 2\pi x_i (\gamma)
\end{equation}
provided that the normalization $\zeta$ is fixed such that conformal invariance
is applicable. We find $\zeta$ from the requirement $x_{\epsilon,s}=1$, which is
also a necessary condition for the marginality of the coupling $\gamma$. 

The conformal theory is in this case more complicated than for
the ordinary or extraordinary transitions. For the spin $\frac{1}{2}$ Ising
model, one can construct the Hamiltonian spectrum either through 
non-unitary Virasoro generators \cite{Henk87}, Kac-Moody 
algebras \cite{Baak89} or alternatively rely
on boundary conformal field theory \cite{Oshi96}. Here we shall restrict 
ourselves to a simple way to characterize the spectrum.  

Taking the spin $\frac{1}{2}$ case as a guide, we expect
that for $n$ large, the low-lying excitation spectrum of $H$ can be 
recovered from  the free fermion Hamiltonian \cite{Henk87,Henk89}
\begin{eqnarray}
H &=& \frac{2\pi}{L} 
\sum_{r=0}^{\infty} \left[ \left( r+\frac{1}{2}-\Delta
\right) n_r^{(-)} + \left( r+\frac{1}{2}+\Delta\right) n_r^{(+)}\right]
\nonumber \\
 &-& \frac{\pi}{6L}\left( \frac{1}{2} - 6\Delta^2 \right) \label{FrFe} 
\end{eqnarray}
where $n_{r}^{(\pm)}$ are fermionic number operators and $\Delta$ depends
on $\gamma$. Non-universal terms which do not enter into the gaps are
already subtracted. 
In general, for states in the even sector (with an even number
of occupied fermionic states) and in the odd sector, there will be 
different values $\Delta_0(\gamma)$ and $\Delta_1(\gamma)$, 
respectively. Now the lowest levels of $H$ can be easily
written down in terms of $\Delta_{0,1}$. For example, the lowest exponents
in the odd sector are
\begin{eqnarray}
x_{\sigma,s}  &=& \frac{1}{2}(\Delta_1 -1)^2 -\frac{1}{2}\Delta_0^2 \nonumber \\
x_{\sigma',s} &=& \frac{1}{2}(\Delta_1 +1)^2 -\frac{1}{2}\Delta_0^2 \label{xung}
\end{eqnarray}
and the
lowest exponents in the even sector are 
\begin{eqnarray} 
x_{\epsilon,s}   &=& 1 \nonumber \\
x_{\epsilon',s}  &=& 2-2\Delta_0  \label{xger}\\
x_{\epsilon'',s} &=& 2+2\Delta_0 \nonumber
\end{eqnarray}
All these exponents correspond to conformal highest weight states. 
In addition, conformal invariance implies that if the exponent $x_i$ 
of a highest weight state occurs in the spectrum,
also $x_i+k$ with $k=1,2,3\ldots$ is present, 
with a known degeneracy which only depends on $k$
(and which is 1 for the lowest two levels). 
From equations (\ref{xung},\ref{xger}), 
the values of $\Delta_{0,1}$ for a given $\gamma$ are found. 
While these are known exactly for the spin $\frac{1}{2}$ case 
\cite{Henk89,Hinr90,Berc91,Zhan96}, these have be determined numerically 
for the case at hand. 

We first fix the normalization constant $\zeta$ from the condition 
$x_{\epsilon,s}=1$. This condition means that the scaled lowest gap 
$A_{even}=n\Delta E_{even}=\frac{1}{2}L \Delta E_{even}$
in the even sector should be equal to $\pi\zeta$, see (\ref{turban}). 
In table~\ref{tab2}, we give $A_{even}$
for several values of $\gamma$. We find that within
our numerical accuracy, its value is independent of $\gamma$ 
(the apparent deviation seen for $\gamma=0.5$ is an artifact from the 
extrapolation of our short sequences and should disappear if larger lattices
could be taken into account) and conclude
that the normalization $\zeta$ is independent of $\gamma$.  
That is only to be expected from earlier results for spin $\frac{1}{2}$ Ising
models with defect lines \cite{Henk89,Hinr90,Berc91}. The final value
of $\zeta$ is taken from the values of $\gamma=0.75\ldots$ and 
$\gamma=0.87\ldots$ where convergence is best and we obtain $\zeta=0.5964(2)$. 

The numerical estimation for the higher gaps is made difficult by 
(a) the relatively short sequences available ($n\leq 7$) and 
(b) the fact that for $n$ finite,
level crossings between different sequences occur.  
In table~\ref{tab3} we give the extrapolated results
for the critical exponents $x_i = L/(2\pi\zeta) (E_i - E_0)$ for several
values of $\gamma$. When no information is given, our sequences did not
converge reliably. We now want to compare these with the spectrum following
from (\ref{FrFe}). First, we use (\ref{xung},\ref{xger}) to determine
$\Delta_{0,1}$, which are also given in table~\ref{tab3}. Depending on the
value of $\gamma$, it turned out to be numerically preferable to
fix first $\Delta_0$ from (\ref{xger}) and than use this value and the
estimate of $x_{\sigma}$ to find $\Delta_1$ or alternatively determine
$\Delta_1$ from the difference $x_{\sigma',s}-x_{\sigma,s}$, 
which is independent
of $\Delta_0$. The values of $\Delta_{0,1}$ were then used to calculate the
other exponents which are listed in table~\ref{tab3} as `expected'. When no
error is given in these columns, the expected value is exact. 

We see that in general the extrapolated estimates for the higher gaps agree 
with the conformal invariance prediction to within a few per cent. 
A particular problem arises
for $\gamma=0.5$, where the converge for $x_{\epsilon}$ is particularly slow.
In that case, we are not able to sensibly to specify accuracies for 
$\Delta_{0,1}$ and the correspondence between the `numerical' and the
`expected' data is more qualitative. The situation here could only be
improved by going to larger lattices. On the other hand, for the other
values of $\gamma$, we obtain a nice agreement between the `numerical'
data and the `expected' free fermion spectrum. We point out that the beginning
of several conformal towers (that is, with $x_i$ also $x_{i}+1$ and even 
$x_{i}+2$ are found in the spectrum) is observed.
The fact that this level spacing comes out correctly is a further confirmation
of our determination of the normalization constant $\zeta$. On the other hand,
we have not been able to go sufficiently high in the spectrum to check the
degeneracies of the excited states. 

We see that the scaling behaviour of our model is described in terms of
a free fermion system. This should be expected on the basis of universality,
although this free fermion description of a spin $1$ model is not at all
obvious from the lattice formulation. Nevertheless, there is an important
distinction with respect to the spin $\frac{1}{2}$ case. Recall that 
the value of $\gamma$ is the same at both subsystem boundaries. Had we coupled
two spin $\frac{1}{2}$ systems, we would have found \cite{Henk89,Hinr90,Berc91}
$\Delta_0=0$, which is not the case in our model, see table~\ref{tab3}. 

\section{Order parameter profiles}

So far, we have calculated the critical exponents which describe the
scaling of observables close to the subsystem boundary. We now ask for the
form of the order parameter profiles close to that interface. 

\subsection{Generalities}

The calculation of the order parameter on a finite lattice poses a conceptual 
problem. The natural candidate, $<M> =\langle 0|M|0\rangle =0$ on any finite
lattice. This difficulty can be overcome by first introducing a small magnetic
field $h$, calculating $<M>$ in the presence of $h$, take the infinite system
limit $L\rightarrow\infty$ and only then let $h\rightarrow 0$. 
In practice, rather than
performing numerically this double limit, the following trick which goes back
to Yang is used. In the ordered phase(s), the ground state is already on a
finite lattice almost degenerate, where the energy splitting decreases 
exponentially with $L$. Introducing an infinitesimal magnetic field $h$ into
$H$ and working within degenerate first-order perturbation theory in $h$, the
order parameter on the site $i$ is given by \cite{Yang52}
\begin{equation} \label{sponM}
m(i) = \langle 1| M(i) |0\rangle
\end{equation}
where $|0\rangle$ and $|1\rangle$ are the lowest eigenstates in the even and odd
sectors, respectively. For our model (\ref{HamDef}), the magnetization
operator $M(i)$ is
\begin{equation}
M(i) = \left\{ \begin{array}{cc} 
\sigma^z_i & \mbox{\rm ; spin $\frac{1}{2}$ region} \\
S^z_i    & \mbox{\rm ; spin $1$ region} \end{array} \right.
\end{equation}
so that $M(i)$ is normalized such that $|M(i)|\leq 1$ for all sites. 
It is well known \cite{Uzel81} that the finite-lattice order parameter
calculated from (\ref{sponM}) has the correct scaling behaviour. The dependence
of the order parameter profiles on $S$ deep in the ordered phase has also been
studied \cite{Henk95}.

Practically,
for the computation of the eigenvectors $|0\rangle, |1\rangle$, it is not
necessary to store all the intermediate L\'anczos vectors. This can
be avoided by running the L\'anczos algorithm twice, where the first pass 
furnishes the weights by which the intermediate vectors contribute to
$|0\rangle, |1\rangle$ and in the second pass the eigenvectors themselves can
be accumulated \cite{Dago94}. 

\subsection{Ordinary and extraordinary transitions}

Before presenting our results for the order parameter profiles at the
various transitions, let us adapt the predictions from finite-size scaling
theory \cite{Barb83,Priv90,Uzel81} to the situation at hand. 
We are interested in the local order
parameter $m(i)$ rather than the full magnetization $m=\sum_i m(i)$. One should
distinguish whether $m(i)$ is measured far away or close to the subsystem
boundary. In the first case, when the site $i$ is well in the bulk, we
expect
\begin{equation} \label{bulkpro}
m(i) = n^{-x_{\sigma}} {\cal M}\left(\frac{2i-1}{4n}\right) 
\end{equation} 
In the second case, when $i$ is close to the boundary, we should have\cite{Skal}
\begin{equation} \label{surfpro}
m(i) = a^{-x_{\sigma}} \left(\frac{n}{a}\right)^{-x_{\sigma,s}} 
\widetilde{\cal M}\left(\frac{i-n-1/2}{a}\right) 
\end{equation}
Here, $x_{\sigma}$ and $x_{\sigma,s}$ are the bulk and 
surface critical exponents
calculated in the previous section, $\cal M$ and 
$\widetilde{\cal M}$ are scaling functions, $i=1,2,\ldots,2n$ is measured from 
the left boundary of the $S=1$ subsystem, $n$ is the layer thickness and $a$ 
the lattice constant. We point out that the arguments of the two scaling 
functions are different.\cite{Skal} In the first case, 
the scaling is such that the total 
system size is kept fixed and the lattice constant $a\rightarrow 0$, 
while in the 
second case, $a$ is kept fixed and the system size $L=2n\rightarrow\infty$. 

\begin{figure}
\centerline{\epsfxsize=3.25in\epsfbox
{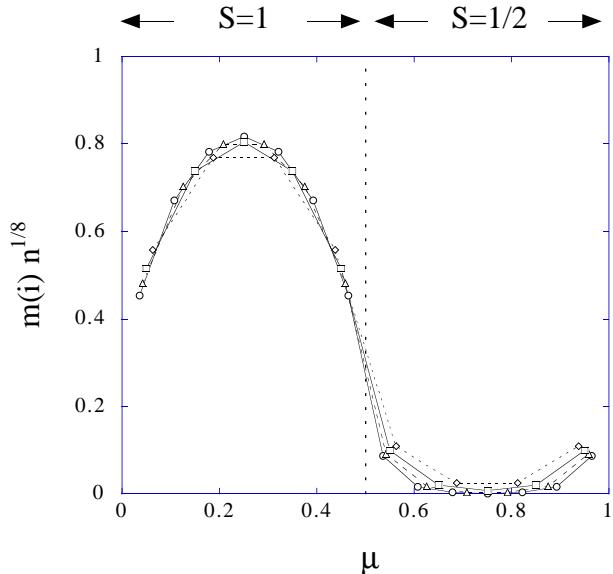}}
\caption[Order parameter bulk]{ 
Order parameter for the ordinary transition, scaled with the bulk exponent
$x_{\sigma}=1/8$ 
for $\kappa=4$ and $\gamma=1$ as a function of $\mu=(2i-1)/4n$. 
The regions of spin $S=1$ and $S=\frac{1}{2}$
are indicated and the boundary between them is shown by the dotted line.
The symbols correspond to $n=4$ (diamonds), $n=5$ (triangles),
$n=6$ (squares) and $n=7$ (circles). \label{fig5}
}
\end{figure}

These predictions are confirmed by our numerical results. Consider first
the ordinary transition. Again, we take $\kappa=4$ and $\gamma=1$ as an
example. In figure~\ref{fig5}, we rescale our magnetization profiles according
to (\ref{bulkpro}) with $x_{\sigma}=1/8$ and we see that indeed for the portion
of the lattice which is far enough for the subsystem boundaries, a data 
collapse occurs even for the small lattices considered here. Also, we see that
in the immediate vicinity of the subsystem boundaries, the scaling
description (\ref{bulkpro}) no longer applies. 
Similar plots for the bulk scaling
can be obtained for the other transitions but will not be presented here,
but see figure~\ref{fig9} below. 

\begin{figure}
\centerline{\epsfxsize=3.25in\epsfbox
{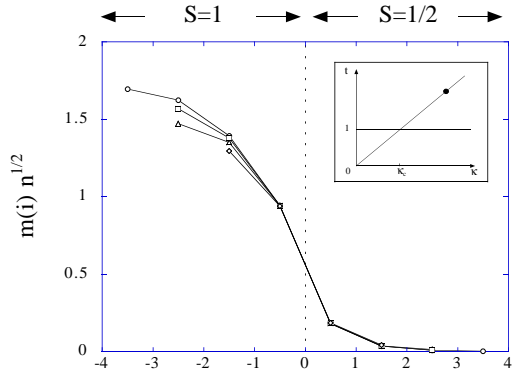}}
\vglue -65mm
\caption[Order parameter for ordinary transition]{Scaled profile 
of the order parameter for the ordinary transition,
for $\kappa=4$ and $\gamma=1$ as a function of 
$\widetilde{\mu}=i-n-\frac{1}{2}$. 
The inset shows the location of the transition point in the
phase diagram. The correspondence of the symbols to the layer thickness
$n$ is the same as in figure~\ref{fig5}. \label{fig6}
}
\end{figure}

In figure~\ref{fig6}, we display the local scaling of the order parameter
close to one of the subsystem boundaries according to (\ref{surfpro}). For
the ordinary transition, $x_{\sigma,s}=1/2$ and we set $a=1$. 
We see that in the first two
monolayers around both sides of the interface, the data collapse onto the
scaling form (\ref{surfpro}). However, going beyond the first two 
monolayers, it is apparent that there is a crossover towards the bulk scaling
form (\ref{bulkpro}). It is apparent that the surface scaling only occurs
in a very thin layer close to the boundary. We remark that this is consistent
with experimental observations that thin magnetic layers on a non-magnetic
substrate show two-dimensional critical behaviour for layer thicknesses
of less than about two monolayers and cross over to three-dimensional
criticality for only slightly thicker layers \cite{Babe87}.

\begin{figure}
\centerline{\epsfxsize=3.25in\epsfbox
{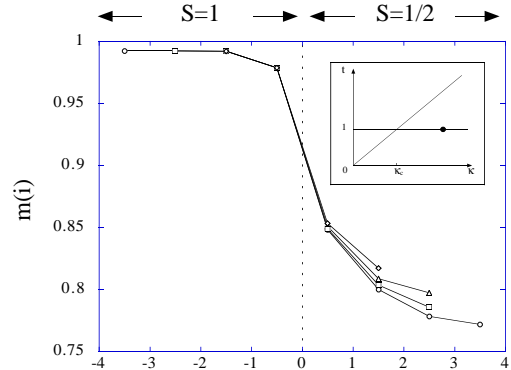}}
\vglue -6.5cm
\caption[Order parameter extraordinary transition]{Profile for
the order parameter $m^{(1)}$ at the extraordinary transition 
for $\kappa=4$ and $\gamma=1$ as a function of $\widetilde{\mu}$. 
The inset shows the location of the transition
point in the phase diagram. The correspondence of the symbols to the layer
thickness $n$ is the same as in figure~\ref{fig5}. \label{fig7}
}
\end{figure}

A similar behaviour is also found for the extraordinary transition. However,
as already mentioned in discussing the spectra, it is sensible to distinguish
two `order parameters'. These are 
\begin{eqnarray}
m^{(1)}(i) &=& \langle 1| M(i) |0\rangle \nonumber \\
m^{(2)}(i) &=& \langle 1| M(i) |0'\rangle \simeq \langle 1'|M(i)|0\rangle
\label{extraM}
\end{eqnarray}
where $|0'\rangle$ and $|1'\rangle$ are the first excited states in the even
and odd sectors and the approximate 
equality between the two forms for $m^{(2)}$
holds up to terms exponentially small in $L$. Note that here the ordering
at the subsystem boundary is provided through the subsystem already in its
ordered phase for $n\rightarrow\infty$ and {\em not} through fixing the spins
at the boundary. The profile for $m^{(1)}$,
where the surface exponent $x_{\sigma,s}^{(1)}=0$, 
is shown in figure~\ref{fig7}. 
Again, we see that for the $S=\frac{1}{2}$ subsystem, we have a data collapse
according to (\ref{surfpro}) for the first two monolayers next to the boundary 
and for larger values of $i$, the is a rapid crossover toward the bulk scaling
(\ref{bulkpro}). Since the $S=1$ subsystem is ordered, finite-size effects
are exponentially small there. 

\subsection{Special transition}

This case is of particular interest, since the exponent $x_{\sigma,s}$ does
depend on $\gamma$. It is therefore interesting to ask whether the profiles
are affected by changing $\gamma$ as well. The bulk scaling  behaviour
(\ref{bulkpro}) with $x_{\sigma}=1/8$ is recovered as in the other transitions.

\begin{figure}
\centerline{\epsfxsize=3.25in\epsfbox
{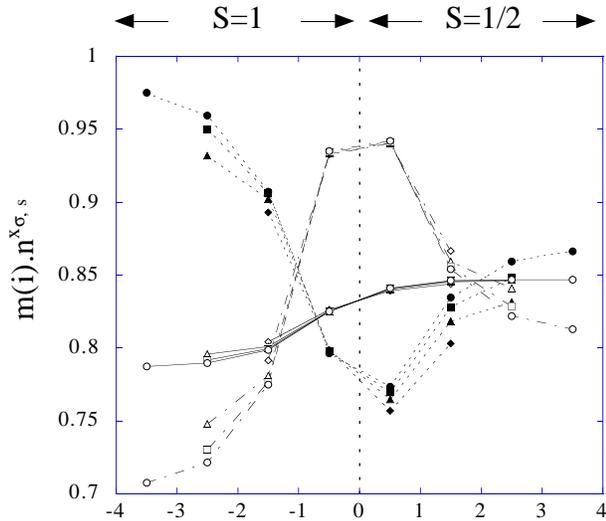}}
\caption[Order parameter special transition]{Scaled profiles for 
the order parameter at the special transition for
several values of $\gamma$ as a function of $\widetilde{\mu}$. 
The dotted curves with full symbols correspond
to $\gamma=\frac{1}{2}$, the full curves with open symbols to $\gamma=0.877111$
and the dash-dotted curves with open symbols corresponds to $\gamma=2$. 
The correspondence of the symbols to the layer thickness $n$ is the same
as in figure~\ref{fig5}. \label{fig8}
}
\end{figure}
In figure~\ref{fig8}, we show for three values of $\gamma$ the local scaling
of the order parameter, where the values of $x_{\sigma,s}$ are taken from
table~\ref{tab3}. On both sides of the boundary, we find a data
collapse according to (\ref{surfpro}) for the first few boundary layers and
for larger values of $i$, a crossover towards the bulk scaling (\ref{bulkpro}). 
In addition, we see that the form of the scaling function $\widetilde{\cal M}$
does depend on $\gamma$. For $\gamma\simeq 0.87$, we have
$x_{\sigma,s}\simeq x_{\sigma}=1/8$ (see table~\ref{tab3}) and the distinction
between local and bulk scaling is somewhat washed out. 

Concerning the shape of the scaling function 
$\widetilde{\cal M}(\widetilde{\mu})$, we see 
that for the special transition, it can be a non-monotonous function of 
$\widetilde{\mu}$. 
For the ordinary and the extraordinary
transitions, however, it is a monotonous function 
of $\widetilde{\mu}$ and this
holds independently of the value of $\gamma$. 
For the special transition, $\widetilde{\cal M}(\widetilde{\mu})$ is only
monotonous if $\gamma\simeq \kappa_c$. Qualitatively, this can be explained
as follows. 

First, if $\gamma\simeq\kappa_c$, the boundary coupling takes the effective
mean value which smoothly interpolates between the two different regimes. 
Since both systems become critical simultaneously at the special transition, 
the scaling functions
simply interpolates smoothly between the values of 
the magnetization finite-size
scaling amplitudes in the two subsystems. Since these amplitudes are
different, even for $x_{\sigma,s}(\gamma)=1/8$, the scaling function
$\widetilde{\cal M}(\widetilde{\mu})$ will not become a constant. 
Second, consider 
$\gamma >\kappa_c$. Then the spins on both sides of the boundary are more
strongly coupled together than two spins in either subsystem. Since the $S=0$
state in the spin $1$ subsystem does not contribute to the energy, this leads
to an enhancement of states where the boundary spins on both sides are up. 
Indeed, we checked that already for $\gamma=4$, the local order parameter
on both sides of the boundary is close to saturation. Thus, we have a large
value of $m(i)$ close to the boundary which then falls back to an average
value for each of the subsystem, in agreement with figure~\ref{fig8}. Finally,
for $\gamma <\kappa_c$, the spins on both sides of the boundary are more weakly
coupled than average spins. This favors states with a smaller value of $m(i)$
close to the subsystem boundary. 

On the other hand, for the ordinary or the extraordinary transition, one 
subsystem is much more ordered than the other one. If $\gamma$ is large, the
first spin across the boundary is strongly aligned with the spins of the
more ordered subsystem and if $\gamma$ is small, the coupling of the first spin
to the more ordered subsystem is reduced. This leads to an effective translation
of the order parameter profile without affecting its form.   

\subsection{Magnetization profiles and conformal invariance} 

In $2D$, conformal invariance states that the profile of a local scaling 
operator $\varphi$ with bulk scaling dimension $x_{\varphi}$ 
is on an infinitely 
long strip of finite width $L$ and with the same type of boundary conditions 
on both sides (and in particular for free boundary conditions) 
given by \cite{Burk85}
\begin{equation} \label{ConfPro}
\langle \varphi(v) \rangle = {\cal A}_{\varphi} \left( \frac{L}{\pi}
\sin\left( \frac{\pi v}{L}\right) \right)^{-x_{\varphi}} 
\sim v^{-x_{\varphi}} \;\; , \;\; v\rightarrow 0
\end{equation}
where $v$ measures the position across the strip and ${\cal A}_{\varphi}$ is 
a non-universal constant. The scaling function for mixed boundary conditions
is also known for minimal conformal theories \cite{Burk91}. This result
only depends on the transformation properties of the 
scaling operator $\varphi$. 
Furthermore, this result carries over to the profiles on quantum chains. 

When we try to apply this to the order parameter at the ordinary transition, 
we should find ${\cal A}_{\sigma}=0$ due to symmetry.   
However, the finite-lattice estimates for $m(i)$ obtained from 
eq.~(\ref{sponM}) above involved an infinitesimal magnetic field $h$, 
which (a) invalidates the above
symmetry argument and (b) leads to a new effective exponent
$x_{\varphi}=x_{\sigma}-x_{\sigma,s}$. This is seen as follows.
From our numerical data, we have found the scaling form (\ref{bulkpro})
\begin{equation} 
m_L(z) = L^{-x_{\sigma}} {\cal M}(\mu) \;\; , \;\; \mu=z/L
\end{equation}
On the other hand, close to the boundary, the order parameter should
show surface finite-size scaling $m_L \sim L^{-x_{\sigma,s}}$. This implies
for the scaling function, e.g. Ref. \onlinecite{Dieh86}
\begin{equation} \label{lowmu}
{\cal M}(\mu) \sim \mu^{ x_{\sigma,s}-x_{\sigma} } \;\; , 
\;\; \mu\rightarrow 0
\end{equation}
from which $x_{\varphi}$ can be identified. 
Eq. (\ref{lowmu}) was also confirmed within the framework of the $\epsilon$ 
expansion\cite{Gomp84} and 
for the Ising quantum chain with an aperiodic modulation generated by the 
Fredholm sequence.\cite{Kare96} For the $2D$ Ising model, we remark that
also in the presence of a small surface magnetic field \cite{Rits96} $h_1$ the
spatial dependence of the magnetization near to the surface scales as 
$z^{3/8}$, which is the same as obtained from eq.~(\ref{lowmu}). 

We now try to extend (\ref{lowmu}), derived for small values of $\mu$ only,
to larger values of $\mu$. If we accept\cite{Iglo96} 
that the estimate eq.~{\ref{sponM})
for the spontaneous magnetization transforms covariantly under conformal
transformations, we can combine eqs.~(\ref{ConfPro},\ref{lowmu}).
Then the exact finite-size scaling function at the ordinary transition
would be\cite{Iglo96}
\begin{equation} \label{SkalForm}
{\cal M}(\mu) = A_{\sigma} 
\left( \sin 2\pi\mu \right)^{ x_{\sigma,s}-x_{\sigma} }
\end{equation}
taking into account that for our model, only the section 
$0\leq \mu\leq \frac{1}{2}$ is actually critical at the ordinary transition
for $\kappa>\kappa_c$. 

\begin{figure}
\centerline{\epsfxsize=3.7in\epsfbox
{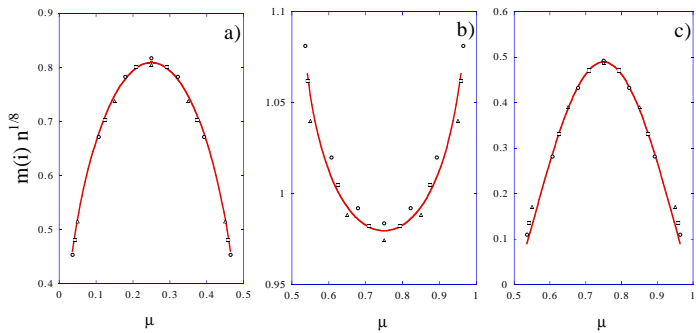}}
\vglue -8.5cm
\caption[Conformal profiles]{Comparison of the conformal invariance prediction
for the scaled order parameter profiles $m(i) n^{1/8}$. Only the part of the
system which is critical is shown. The graphs give 
for (a) the ordinary transition, (b) the 
extraordinary transition with $m^{(1)}$ and (c) for $m^{(2)}$ 
at $\kappa=4$ and $\gamma=1$.
The correspondence of the symbols to the layer thickness $n$ is the same
as in figure~\ref{fig5}. \label{fig9}
}
\end{figure}

In figure~\ref{fig9}a, we compare (\ref{SkalForm}) to the numerical data. 
First, we observe a data collapse from several system sizes onto a single
curve. Second, the form of the scaling function agrees nicely with 
eq.~(\ref{SkalForm}).  
The same result has also been found for the Hilhorst-van Leeuven
model \cite{Iglo96,Kare96}. 
Since for the $2D$ Ising model, (\ref{lowmu}) remains
unchanged even in the presence of a small surface magnetic field,\cite{Rits96}
$\cal M$ should be independent of a
small $h_1$ for the ordinary transition. 

Let us compare the finite-size scaling functions for the profiles coming
from (\ref{ConfPro}) and (\ref{SkalForm}). The first one is based on a 
continuum description of the profile in the half-infinite system which is then 
conformally transformed onto the strip \cite{Burk85}. 
Very close to the boundary, 
a continuum description may no longer be applicable. Indeed, for unitary
conformal theories such as the Ising model, 
the critical exponents $x_{\varphi}>0$ and the profile as it stands will
diverge at the boundary, in disagreement with existing numerical data. 
On the other hand, the second form is constructed to be consistent with both
bulk finite-size scaling deep inside the system and surface finite-size
scaling close to the boundary. In order to match this with the functional
form required from conformal invariance, it is necessary to assume that
the exponent $x_{\sigma}-x_{\sigma,s}$ governs the scaling of the matrix
element (\ref{sponM}) (calculated in an infinitesimal magnetic field
which breaks global symmetry) used to estimate the finite-size order
parameter, rather than the conventional order parameter 
scaling dimension $x_{\sigma}>0$. While this approach certainly
is in agreement with the numerical data for the whole strip, it is not
yet understood how the above dimensional transmutation can be explained. 

In addition, we find that the same functional form also describes the 
order parameter profiles for the extraordinary transition, as shown in
figure~\ref{fig9}. The numerical data are again consistent with
scaling (note that the overall scale in figure~\ref{fig9}b is abount an
order of magnitude larger than in the other two cases). 
Specifically, we find from a fit
\begin{equation}
{\cal M}(\mu) = \left\{ \begin{array}{lll}
A_{\sigma} (\sin 2\pi\mu)^{3/8} & ,\;A_\sigma\;\; \simeq 0.80 
& \mbox{\rm ord.} \\
A_{\sigma}^{(1)} (\sin 2\pi\mu)^{-1/16} & ,\;A_{\sigma}^{(1)} \simeq 0.98 &
\mbox{\rm ex., $m^{(1)}$} \\
A_{\sigma}^{(2)} (\sin 2\pi\mu )^{9/8} & ,\;A_{\sigma}^{(2)} \simeq 0.49 &
\mbox{\rm ex., $m^{(2)}$} 
\end{array} \right.
\end{equation}

Tentatively, the profile for $m^{(1)}$ (which is {\em not} given by 
(\ref{ConfPro}) because the `ordered' subsystem is still finite, see 
eq.~(\ref{extraM})) can be explained as follows. This
order parameter is sensible to the ordering which occurs at the ordinary 
transition. At the extraordinary transition, these degrees of freedom have
become massive and thus have a short effective correlation length $\xi_{eff}$. 
Then the fluctuating spins would see a fixed boundary on one side but because
$\xi_{eff} \ll L$ the other boundary should appear as open. However, for mixed
boundary conditions, it is known that \cite{Burk87,Chri93} 
$x_{\sigma,s}=\frac{1}{16}$. 
Then $-x_{\varphi}=\frac{1}{16}-\frac{1}{8}=-\frac{1}{16}$ in agreement with
the numerical data.

Finally, the above argument does not reproduce the profiles for the special
transition. This is due to the fact that the two-point correlation
functions are more complicated \cite{Oshi96} than the simple power-law
form which underlies the derivation \cite{Burk85} of (\ref{ConfPro}).

\section{Summary}

We have studied the transitions arising in a pair of magnetic layers,
coupled through short-ranged interactions and described by Ising models. 
This study was motivated by ongoing experiments on similar systems. We have
found the variation of the `specific heat' with the temperature and studied the
scaling of the order parameter profiles at the phase transition points. 
Our aim was to check out a scaling analysis which should also be applicable to
experimental data in $3D$. 

We have reemphasized that the systems usually studied in
experiments and the models best suited for theoretical analysis are not
completely identical and care is needed in the comparison of the two, as
exemplified in the two phase diagrams in figure~\ref{fig2}. 

To simplify the theoretical analysis, we used a two-dimensional model of
layers of spin chains, although the experiments \cite{Lede93,Babe87,Baue95} 
involve two-dimensional layers stacked in the third dimension. This was for
us no serious disadvantage, since the scaling arguments used here can be
extended to three dimensions in a well-known way \cite{Lede93a,Barb83,Priv90}.
In addition, conformal invariance techniques could be used to simplify the
calculation of the surface critical exponents we were interested in. 
This calculation reconfirmed the expected universality of the surface
critical exponents found for the ordinary, extraordinary and special
transition (where in the $2D$ Ising model, the surface coupling is marginal). 

Our results on how the `specific heat' depends on the layer thickness $n$ are
qualitatively the same as seen experimentally \cite{Lede93}. Our results
also suggest that if the layer thickness can be precisely controlled, one
might get a trade-off in no longer having to achieve a very fine temperature
control and still being able to measure fluctuation-dominated critical
exponents to good accuracy. 

Our study for the order parameter profiles was motivated by the existing
experimental techniques to measure the magnetic moments of a single monolayer
\cite{Baue95}. While for an infinite system, the critical point magnetization
vanishes, for finite lattices a non-trivial finite-size scaling behaviour
of the order parameter profiles is obtained. 
We found two types of finite-size scaling
at the transitions. Far away from the subsystem boundaries, we get a bulk
finite-size scaling (\ref{bulkpro}) 
governed by the bulk exponent $x_{\sigma}=\beta/\nu=1/8$.
In the immediate vicinity (and on {\em both} sides) of the boundary, 
however, the order parameter
finite-size scaling (\ref{surfpro}) is governed by the 
surface critical exponent $x_{\sigma,s}=\beta_1/\nu$ whose value depends 
on the type of the surface transition. 

For the ordinary and the extraordinary transitions, the data for 
finite-size scaling
of the order parameter profile scaling function $\cal M$ (measured in an
infinitesimal magnetic field) suggest in $2D$ an exact scaling function
from consistency with bulk and surface finite-size scaling and
with conformal invariance. It remains a challenge to derive
a similar result for the special transition. It is not yet clear how to 
address the problem of calculating the surface scaling function 
$\widetilde{\cal M}$. In any case, a continuum approach, 
which underlies the conformal 
invariance arguments used for the determination of $\cal M$, 
does not seem to be feasible in that case.  
~ \\~ \\  
It is a pleasure to thank R. Camley, F. Igl\'oi and L. Turban 
for useful discussions. 



\newpage ~
\newpage 
\widetext

\begin{table}
\begin{tabular}{cccccc}
 &\multicolumn{2}{c}{ordinary}& & \multicolumn{2}{c}{extraordinary}\\ \hline
$i$ & numerical & expected & & numerical & expected \\ \hline
0 & 0         & 0     & & 0       & 0  \\
1 & 0.4994(6) & $1/2$ & & 2       & 2 \\
2 & 1.496(7)  & $3/2$ & & 3.01(2) & 3 \\
3 & 2         & 2     & & 3.95(6) & 4 \\
4 & 2.505(9)  & $5/2$ & & 4.9(2)  & 5 \\
\end{tabular}
\caption{Conformal spectrum of the surface exponents $x_i$
for the ordinary and extraordinary transitions
at $\kappa=4$ and $\gamma=1$. The numbers in brackets give the estimated
uncertainty in the last given digit. \label{tab1}}
\end{table}

\begin{table}
\begin{tabular}{clllll}
$\gamma$   & 0.5      & 0.65     & 0.754222 & 0.877111 & 1 \\
$A_{even}$ & 1.945(3) & 1.889(7) & 1.8731(8) & 1.8745(5) & 1.88(2) \\
\end{tabular}
\caption[energy gaps]{Scaled and extrapolated lowest gap 
$A_{even}=\pi\zeta$ in the even sector at the special transition for 
several values of $\gamma$. The numbers in brackets give the estimated 
extrapolation error in the last digit. \label{tab2}}
\end{table} 

\begin{table}
\begin{tabular}{cllllllll}
 & \multicolumn{2}{c}{$\gamma=0.5$} & \multicolumn{2}{c}{$\gamma=0.754222$} &
 \multicolumn{2}{c}{$\gamma=0.877111$} & \multicolumn{2}{c}{$\gamma=1$} \\
$i$ & numerical & expected & numerical & expected 
    & numerical & expected & numerical & expected \\ \hline
1 & 0.231(1) & 0.20 & 0.1436(6) & 0.144(2) & 0.1103(7) & 0.1103(5) 
& 0.0841(5) & 0.084(5) \\
2 & 0.971(3) & 0.91 & 0.999(1)  & 1        & 1.000(1)  & 1         
& 1.00(2)   & 1 \\
3 & 1.034(5) & 1    & 1.10(3)   & 1.072(4) & 1.113(6)  & 1.1103(5) 
& 1.11(2)   & 1.084(5) \\
4 & 1.095(3) & 1.20 & 1.12(2)   & 1.144(2) & 1.168(5)  & 1.168(2)  
& 1.232(5)  & 1.232(5) \\
5 & 1.72(3)  & 1.70 & 1.94(1)   & 1.94(1)  & 1.92(1)   & 1.92(1)   
& 1.9(1)    & 1.78(4) \\
6 & 1.92(3)  & 1.91 & 1.995(5)  & 2        & 2.00(2)   & 2         
& 2.0(1)    & 2 \\
7 & 1.99(2)  & 2    & 2.03(3)   & 2.06(1)  & 2.11(1)   & 2.08(1)   
& 2.1(1)    & 2.084(5) \\
8 & -        & 2.20 & 2.11      & 2.072(4) & -         & 2.1103(5) 
& 2.18(3)   & 2.22(4) \\
9 & -        & 2.30 & 2.145(8)  & 2.144(2) & -         & 2.168(2)  
& -         & 2.232(5) \\ \hline
$\Delta_0$ & \multicolumn{2}{c}{0.15} & \multicolumn{2}{c}{0.030(5)} & 
\multicolumn{2}{c}{0.040(5)} & \multicolumn{2}{c}{0.11(3)~ } \\
$\Delta_1$ & \multicolumn{2}{c}{0.36} & \multicolumn{2}{c}{0.464(2)} &
\multicolumn{2}{c}{0.529(1)} & \multicolumn{2}{c}{0.574(4)} \\ 
\end{tabular}
\caption[scaling dimensions]{Conformal spectrum of the 
scaling dimensions $x_i(\gamma)$ at the
special point $t=1, \kappa=0.754222$. The values of $\Delta_{0,1}$ 
used in comparing
with the free fermion Hamiltonian (\ref{FrFe}) are also given. The numbers
in brackets give the estimated uncertainty in the last given digit. 
\label{tab3}}
\end{table}


\begin{references}
\bibitem[*]{URA} Unit\'e de recherche associ\'ee au CRNS no. 155
\bibitem{Lede93a} D. Lederman, C.A. Ramos, V. Jaccarino and J.L. Cardy, Phys.
Rev. {\bf B48}, 8365 (1993).
\bibitem{Barb83} M.N. Barber in Domb and Lebowitz (Eds) {\it Phase Transitions
and Critical Phenomena}, Vol. 8, ch. 2, Academic Press (New York 1983).
\bibitem{Priv90} V. Privman (Ed) {\it Finite Size Scaling and Numerical 
Simulation of Statistical Systems}, World Scientific (Singapore 1990). 
\bibitem{Lede93} D. Lederman, C.A. Ramos and V. Jaccarino, J. Phys. Cond. 
Matt. {\bf 5}, A373 (1993).
\bibitem{Baue95} Ph. Bauer, S. Andrieu and M. Piecuch, Nouv. Cim. {\bf 18D}, 
299 (1996); Ph. Bauer, S. Andrieu, M. Lemine and M. Piecuch, E-MRS Conference,
Stra{\ss}burg June 1996, to appear in J. Mag. Mag. Mat. 
\bibitem{Suzu71} M. Suzuki, Prog. Theor. Phys. {\bf 46}, 1337 (1971); 
E. Fradkin and L. Susskind, Phys. Rev. {\bf D17}, 2637 (1978).  
\bibitem{Chri93} P. Christe and M. Henkel, {\em Introduction to Conformal 
Invariance and its Applications to Critical Phenomena}, Springer (Berlin 1993).
\bibitem{Bitk96} D. Bitko, T.F. Rosenbaum and G. Aeppli, Phys. Rev. Lett. 
{\bf 77}, 940 (1996). 
\bibitem{Hinr90} H. Hinrichsen, Nucl. Phys. {\bf B336}, 377 (1990).
\bibitem{Berc91} B. Berche and L. Turban, J. Phys. {\bf A24}, 245 (1991).
\bibitem{Zhan96} D-G. Zhang, B.-Z. Li and M.-G. Zhao, Phys. Rev. {\bf B53}, 
8161 (1996).
\bibitem{Pfeu70} P. Pfeuty, Ann. of Phys. {\bf 57}, 79 (1970).
\bibitem{Hofs96} W. Hofstetter and M. Henkel, J. Phys. {\bf A29}, 1359 (1996).
\bibitem{Dago94} E. Dagotto, Rev. Mod. Phys. {\bf 66}, 763 (1994).
\bibitem{alnull} If $\alpha=0$ and $C$ has a logarithmic singularity 
as happens for the $2D$ Ising model, a similar
analysis shows that $C \sim \ln \xi_2$.
\bibitem{xis} In the simple spin systems usually considered this distinction
is not necessary and $\xi_1$ and $\xi_2$ are propotional to each other.
\bibitem{Ramo90} C.A. Ramos, D. Lederman, A.R. King and V. Jaccarino, 
Phys. Rev. Lett. {\bf 65}, 2913 (1990).
\bibitem{Dieh86} H.W. Diehl in Domb and Lebowitz (Eds) {\it Phase Transitions
and Critical Phenomena}, Vol. 10, ch. 2, Academic Press (New York 1986). 
\bibitem{zweidrei} Since the model eq.~(\ref{HamDef}) is two-dimensional, 
the layers cannot order for $n$ finite and thus there is no 
{\em surface} transition. This would be different in a three-dimensional model,
 where surface transitions may occur \cite{Dieh86}.
\bibitem{Card87} J.L. Cardy, in Domb and Lebowitz (Eds) {\it Phase Transitions
and Critical Phenomena}, Vol. 11, ch. 2, Academic Press (New York 1987). 
\bibitem{Itzy89} C. Itzykson and J.-M. Drouffe, {\it Statistical Field Theory},
Vol. 2, ch. 9, Cambridge University Press (Cambridge 1989). 
\bibitem{Iglo93} F. Igl\'oi, I. Peschel and L. Turban, Adv. Phys. {\bf 42}, 
683 (1993).
\bibitem{Card84} J.L. Cardy, J. Phys. {\bf A17}, L385 (1984). 
\bibitem{Burk87} T.W. Burkhardt and I. Guim, Phys. Rev. {\bf B35}, 1799 (1987);
J.L. Cardy, Nucl. Phys. {\bf B275}, 200 (1986); G.v. Gehlen and V. Rittenberg,
J. Phys. {\bf A19}, L631 (1986). 
\bibitem{Henk88} M. Henkel and G. Sch\"utz, J. Phys. {\bf A21}, 2617 (1988).
\bibitem{norm} These values for $\zeta$ are close to the ones found for the
spin $\frac{1}{2}$ and spin $1$ Ising model separately \cite{Hofs96}, thereby
confirming that only the critical degrees of freedom make a contribution 
to $\zeta$.
\bibitem{Bray77} A.J. Bray and M.A. Moore, J. Phys. {\bf A10}, 1927 (1977). 
\bibitem{Bari79} R.V. Bariev, Sov. Phys. JETP {\bf 50}, 613 (1979);
B.M. McCoy and J.H.H. Perk, Phys. Rev. Lett. {\bf 44}, 840 (1980);
L.P. Kadanoff, Phys. Rev. {\bf B24}, 5382 (1981);
D.B. Abraham, L.F. Ko and N.M. \v Svrakic, J. Stat. Phys. {\bf 56}, 563 (1989).
\bibitem{Turb85} L. Turban, J. Phys. {\bf A18}, L325 (1985). 
\bibitem{Henk87} M. Henkel and A. Patk\'os, Nucl. Phys. {\bf B285}, 29 (1987).
\bibitem{Henk89} M. Henkel, A. Patk\'os and M. Schlottmann, Nucl. Phys. {\bf
B314}, 609 (1989).
\bibitem{Oshi96} G. Delfino, G. Mussardo and P. Simonetti, Nucl. Phys. 
{\bf B432}, 518 (1994); M. Oshikawa and I. Affleck, Vancouver preprint, 
hep-th/9606177.
\bibitem{Baak89} M. Baake, P.Chaselon and M. Schlottmann, Nucl. Phys. 
{\bf B314}, 625 (1989). 
\bibitem{Yang52} C.N. Yang, Phys. Rev. {\bf 85}, 808 (1952).
\bibitem{Uzel81} K. Uzelac and R. Jullien, J. Phys. {\bf A14}, L151 (1981);
C.J. Hamer, J. Phys. {\bf A15}, L675 (1982).
\bibitem{Henk95} H.J. Mikeska, S. Miyashita and G.H. Ristow, J. Phys. Cond.
Matt. {\bf 3}, 2985 (1991); M. Henkel, A.B. Harris and M. Cieplak, Phys. Rev. 
{\bf B52}, 4371 (1995).
\bibitem{Skal} These forms can be obtained from $m_L(z)=L^{-x_{\sigma}}
\Phi(z/L,L/a)$ where $z$ is the distance across the strip. Eq. (\ref{bulkpro})
is recovered in the limit $a\rightarrow 0$. Eq. (\ref{surfpro}) is obtained
in the limit $L\rightarrow\infty$ with $a$ fixed 
and $\Phi(u,v)\simeq u^{\omega_1}
v^{\omega_2}\widetilde{\Phi}(uv)$. 
\bibitem{Babe87} M. Farle and K. Baberschke, Phys. Rev. Lett. {\bf 58}, 511 
(1987); W. D\"urr, D. Kerkmann and D. Pescia, Int. J. Mod. Phys. {\bf B4}, 401
(1990); Z.Q. Qiu, J. Pearson and S.D. Bader, Phys. Rev. Lett. {\bf 67}, 1646
(1991); C. Rau and C. Jin, J. Physique Colloque C8, {\bf 49}, C8-1627 (1988). 
\bibitem{Burk85} T.W. Burkhardt and E. Eisenriegler, J. Phys. {\bf A18}, L83
(1985).
\bibitem{Burk91} T.W. Burkhardt and T. Xue, Nucl. Phys. {\bf B354}, 653 (1991).
\bibitem{Gomp84} G. Gompper, Z. Phys. {\bf B56}, 217 (1984).
\bibitem{Kare96} D. Karevski, th\`ese de doctorat, Nancy 1996. 
\bibitem{Rits96} U. Ritschel and P. Czerner, Essen preprint, cond-mat/9603011. 
\bibitem{Iglo96} F. Igl\'oi and L. Turban, to be published. 
\end{references}
\end{document}